\documentstyle[pra,preprint,psfig,aps]{revtex}

\begin{document}
\draft
\title{Quantum Nondemolition State Measurement via Atomic Scattering in Bragg Regime}
\author{Aeysha Khalique and Farhan Saif}
\address{Department of Electronics, Quaid-i-Azam university, 45320, Islamabad,\\
Pakistan.}
\maketitle

\begin{abstract}
We suggest a quantum nondemolition scheme to measure a quantized cavity
field state using scattering of atoms in general Bragg regime. Our work
extends the QND measurement of a cavity field from Fock state, based on
first order Bragg deflection \cite{Arbab}, to any quantum state based on
Bragg deflection of arbitrary order. In addition a set of experimental
parameters is provided to perform the experiment within the frame work of
the presently available technology.
\end{abstract}


\newpage

\section{Introduction}

In twentieth century physics, quantum theory has been a fascinating
playground to study the nature of electromagnetic radiations and matter. In
this subject, measurement of an unknown state of electromagnetic field poses
an interesting question \cite{Ulf}. After the breakthrough made by
tomographic technique \cite{tomo}, numerous methods such as quantum
non-demolition \cite{Zag}, quantum endoscopy \cite{endoscopy} and Autler
Townes spectroscopy \cite{Zub}, have been suggested. In quantum non
demolition (QND) \cite{QND} measurement schemes, field photon number is not
altered by measurement process, which includes QND\ measurement via
dispersive atom field coupling \cite{Zag}, optical Kerr effect \cite{Kerr}
and atomic scattering \cite{Zoller,Arbab}.

Since, momentum distribution of deflected atoms from a quantized cavity
field is a function of photon number \cite{Mystre} of the field, therefore,
atomic scattering provides various methods to measure field photon
statistics. Two such approaches have been employed: $(i)$ Resonant
interactions of atoms with a quantized cavity standing light field \cite
{Herkomer,Frey}; $\left( ii\right) $ Nonresonant interaction of atoms with
the quantized field, which are qantum nondemolition (QND) schemes. In QND
state measurement schemes we reduce field statistics to a single Fock state,
and then make repeated measurements to reproduce the field statistics \cite
{Zag,Zag2,Ueda}. Such schemes have been discussed to measure the field
photon number using a phase sensitive detection of atom \cite{Zag} and
atomic scattering in Raman-Nath regime \cite{Zoller}. Atomic scattering
using first order of Bragg regime has been employed to measure Fock state 
\cite{Arbab} of a cavity field.

In this paper, we study deflection of atoms, in any arbitrary order of Bragg
regime, from a far detuned quantized cavity field. We apply this study to
measure any quantum state using QND state measurement scheme. As an
application of this method, we measure coherent state of electromagnetic
field in a very high Q-cavity. In addition a set of experimental parameters
is provided to perform the experiment within the frame work of the presently
available technology.

In Sec.~2, we present experimental model of our system. In Sec.~3, we
develop mathematical model of our system describing the dynamics of atom in
a quantized cavity field. We present analytical prescription to study the
momentum probability distributions for generalized Bragg regime and provide
numerical confirmation of the derived results. Finally, in Sec.~4, we
present our scheme to measure the quantum state of cavity field and
successfully apply it to reconstruct coherent state. We also provide set of
experimental parameters to realize the suggested scheme.

\section{The Model}

We consider a beam of atoms moving with centre of mass momentum $P,$ towards
a cavity field. The incident atom is in the ground state and the field
inside the cavity is an optical standing wave of wavelength, $\lambda .$ We
have the injection rate to be small enough so that only one atom is in the
cavity at a time. Moreover, we take large detuning between the cavity field
frequency and the atomic transition frequency. This ensures that atom does
not exit the cavity in excited state and there is no spontaneous emission
which contributes a photon in arbitrary direction.

The effective Rabi frequency of the atom becomes $|g|^{2}n/2\triangle =\chi
n,$ where $g$ is coupling constant, and $\triangle =\nu -w_{0}$ is detuning
between the field frequency, $\nu ,$ and the atomic transition frequency, $%
w_{0}.$ We probe the atomic scattering by taking the propagation direction
at an angle $\theta $ to the normal, satisfying the Fresnel approximation.
In this approximation the momentum component, $P_{z},$ normal to the cavity
field, is taken very large compared to the component, $P_{0},$ along the
cavity field, as shown in Fig.~1. This allows us to treat $P_{z}$
classically.

\section{Atomic Scattering from Cavity Field}

We describe the wave function of our system at any time, $t,$ in discrete
momentum space as

\begin{equation}
|\Psi (t)\rangle =\sum_{n}\sum_{l}\left( C_{P_{l}}^{a,n}(t)|a,n,P_{l}\rangle
+C_{P_{l}}^{b,n}(t)|b,n,P_{l}\rangle \right) .  \label{Psi}
\end{equation}
Here, $C_{P_{l}}^{i,n}$ is the probability amplitude indicating the atom in
state $i=a,b$ at any interaction time $t,$ exiting with momentum $P_{l},$
and the cavity with $n$ photons$.$ The total Hamiltonian of the system, in
dipole and in rotating wave approximations, is 
\begin{equation}
H=\frac{P^{2}}{2M}+\hbar \nu a^{\dagger }a+\hbar w_{0}\sigma _{z}+\hbar \cos
kx(g\sigma _{+}a+g^{*}a^{\dagger }\sigma _{-}).  \label{H}
\end{equation}
Since, in presence of large detuning, the probability of finding the atom in
excited state is very small, we may express the atomic dynamics by means of
the effective Hamiltonian as 
\begin{equation}
H_{eff}=\frac{P_{x}^{2}}{2M}-\frac{\hbar |g|^{2}}{2\triangle }\stackrel{%
\symbol{94}}{n}\sigma _{-}\sigma _{+}\left( \cos 2kx+1\right) ,  \label{Veff}
\end{equation}
obtained by applying adiabatic approximation \cite{Scully}. During
interaction with the field, for each complete Rabi cycle, momentum
transferred to the atom by the field is either $zero$ or $2\hbar k$ \cite
{Shore}$.$ Thus momentum of the exiting atom is given as $P_{l}=P_{0}+l\hbar
k$, where, $l$ is an even integer. Hence, Eq. $\left( \ref{Veff}\right) $
leads to a set of infinite coupled rate equations for probability amplitudes 
$C_{P_{l}}^{b,n}$, viz., 
\begin{equation}
i\frac{\partial C_{P_{l}}^{b,n}(t)}{\partial t}%
=w_{rec}l(l+l_{0})C_{P_{l}}^{b,n}(t)-\frac{\chi n}{2}\left( C_{P_{l}+2\hbar
k}^{b,n}(t)+C_{P_{l}-2\hbar k}^{b,n}(t)\right) .  \label{cbf}
\end{equation}
Here, $w_{rec}=\frac{\hbar k^{2}}{2M}$ is the recoil frequency of the atom
and its initial momentum is taken to be, $P_{0}=l_{0}\hbar k.$
Interestingly, Eq. $\left( \ref{cbf}\right) $ helps us to study atomic
scattering in any regime.

In Bragg deflection, recoil frequency of the deflected atom is much larger
than effective Rabi frequency \cite{Arbab,Marte}, that is $w_{rec}\gg \chi n.
$ By comparing atomic scattering with optical Bragg scattering \cite
{Braggorg}, we \cite{Mystre,Shore,Marte,Rempe,Prichard,Pichard,Gould} can
develop a condition on initial momentum of the incident atom, viz. $P_{0}=%
\frac{l_{0}}{2}\hbar k.$ Here, $l_{o}=2,4$, $6$ etc., correspond to first,
second, third order of Bragg scattering, respectively. By changing the
component of momentum, $P_{0},$ parallel to the cavity, we can change the
order of Bragg scattering. In this regime, the conservation of energy
provides us $l=0\ $and\ $l=-l_{0},$ which indicate only two possible
directions of propagation for the scattered atom, one with initial momentum $%
P_{0}$ and the other with momentum $-P_{0},$ respectively, as shown in
Fig.~1.

In presence of above mentioned conditions, Eq. $\left( \ref{cbf}\right) $
yields a set of coupled rate equations from $l=0$ to $l=-l_{0}$ for Bragg
regime, viz.

\begin{eqnarray}
i\frac{\partial C_{P_{0}}^{b,n}}{\partial t} &=&-\frac{\chi n}{2}(\underline{%
C_{P_{2}}^{b,n}}+C_{P_{-2}}^{b,n}),  \label{0} \\
i\underline{\frac{\partial C_{P_{-2}}^{b,n}}{\partial t}}
&=&w_{rec}(-2)(-2+l_{0})C_{P_{-2}}^{b,n}-\frac{\chi n}{2}%
(C_{P_{0}}^{b,n}+C_{P_{-4}}^{b,n}), \\
i\underline{\frac{\partial C_{P_{-4}}^{b,n}}{\partial t}}
&=&w_{rec}(-4)(-4+l_{0})C_{P_{-4}}^{b,n}-\frac{\chi n}{2}%
(C_{P_{-2}}^{b,n}+C_{P_{-6}}^{b,n}), \\
&&\vdots   \nonumber \\
i\underline{\frac{\partial C_{P_{-l_{0}+4}}^{b,n}}{\partial t}}
&=&w_{rec}(-l_{0}+4)(4)C_{P_{-l_{0}+4}}^{b,n}-\frac{\chi n}{2}%
(C_{P_{-l_{0}+6}}^{b,n}+C_{P_{-l_{0}+2}}^{b,n}), \\
i\underline{\frac{\partial C_{P_{-l_{0}+2}}^{b,n}}{\partial t}}
&=&w_{rec}(-l_{0}+2)(2)C_{P_{-l_{0}+2}}^{b,n}-\frac{\chi n}{2}%
(C_{P_{-l_{0}+4}}^{b,n}+C_{P_{-l_{0}}}^{b,n}), \\
i\frac{\partial C_{P_{-l_{0}}}^{b,n}}{\partial t} &=&-\frac{\chi n}{2}%
(C_{P_{-l_{0}+2}}^{b,n}+\underline{C_{P_{-l_{0}-2}}^{b,n}}).  \label{L0}
\end{eqnarray}
At $l=0$ and at $l=-l_{0},$ the diagonal terms of above set of coupled
equations vanish$.$ In the Bragg limit, $w_{rec}$ is much larger than $\chi n
$, the non-vanishing diagonal elements dominate the evolution. Hence, we
ignore, adiabatically, the time derivatives of these probability amplitudes.
Also the probability amplitudes outside this range acquire very little
probability. Thus, we ignore all the underlined terms in above series. Now
retaining only lowest order of $\chi n/2$ in the coefficients and back
substituting the values, we obtain two coupled equations (for $l_{0}>2$) as 
\begin{eqnarray}
i\frac{\partial C_{P_{0}}^{b,n}}{\partial t} &=&AC_{P_{0}}^{b,n}-\frac{1}{2}%
BC_{P_{-l_{0}}}^{b,n},  \label{r1} \\
i\frac{\partial C_{P_{-l_{0}}}^{b,n}}{\partial t} &=&AC_{P_{-l_{0}}}^{b,n}-%
\frac{1}{2}BC_{P_{0}}^{b,n},  \label{r2}
\end{eqnarray}
where, 
\[
A=-\frac{\chi n/2}{w_{rec}(l_{0}-2)(2)},
\]
and 
\begin{equation}
|B|=\frac{\left( \chi n\right) ^{^{\frac{l_{0}}{2}}}}{\left( 2w_{rec}\right)
^{\frac{l_{0}}{2}-1}\left[ (l_{0}-2)(l_{0}-4)......4.2\right] ^{2}}.
\label{b}
\end{equation}
From Eqs. $\left( \ref{r1}\right) $ and $\left( \ref{r2}\right) $, we obtain
the probability of atom exiting with momentum $P_{0},$ i.e. $Q(P_{0},t),$
and that of exiting with momentum $P_{-l_{0}},$ i.e. $Q(P_{-l_{0}},t),$ as 
\begin{eqnarray}
Q(P_{0},t) &=&\sum_{n}P(n)\cos ^{2}\left( \frac{1}{2}Bt\right) ,
\label{Bragg4} \\
Q(P_{-l_{0}},t) &=&\sum_{n}P(n)\sin ^{2}\left( \frac{1}{2}Bt\right) ,
\label{Bragg42}
\end{eqnarray}
where, $P(n)$ is photon statistics. Here, we have used the initial condition
that atom enters the cavity with momentum $P_{0},$ therefore, $%
C_{P_{0}}^{b,n}(0)=1$ and $C_{P_{-l_{0}}}^{b,n}(0)=0.$ Thus, we find that
the probability of finding the exiting atom in either of the two directions,
flips as a function of interaction time $t,$ with frequency, $|B|$.

In the simplest case of first order Bragg scattering, Eq. $\left( \ref{cbf}%
\right) $ provides only two coupled rate equations, that is for $l=0$ and $%
l=-2.$ This leads to the results obtained in Ref. \cite{Arbab}, 
\begin{eqnarray}
Q(P_{0},t) &=&\sum_{n}P(n)\cos ^{2}\left( \frac{\chi n}{2}t\right) ,
\label{Bragg} \\
Q(P_{-2},t) &=&\sum_{n}P(n)\sin ^{2}\left( \frac{\chi n}{2}t\right) .
\label{Bragg2}
\end{eqnarray}

In order to verify our above mentioned analytical results obtained by means
of adiabatic approximation and in presence of Bragg condition, we exactly
solve a system of equations, numerically. The system of first order linear
coupled differential equations is obtained from Eq. $\left( \ref{cbf}\right) 
$ from $l=-500$ to $l=+500.$ We solve this set for Bragg regime and present
the comparison for 2nd order Bragg scattering, that is for $l_{0}=4$, in
Fig.~2. We find a good qualitative and quantitative agreement between
analytical and numerical results.

\section{QND\ Cavity\ State\ Measurement}

QND measurement process relies on the reduction of field statistics to a
single Fock state by means of successive atomic interactions and then
repeating the reduction process, such that, the statistics of the Fock
states reproduces field statistics. From Eqs. $\left( \ref{Bragg4}\right) $
and $\left( \ref{Bragg42}\right) ,$ we see that momentum distribution of the
scattered atoms depends on the photon statistics inside the cavity. If
cavity life time is large compared to the time between injections of atoms,
then it is possible to invert Eq. $\left( \ref{Bragg4}\right) $ or $\left( 
\ref{Bragg42}\right) $. Interaction of an atom with the cavity field updates
the cavity field statistics in a fashion controlled by its interaction time.
The photon statistics, $P(n),$ gets multiplied by oscillatory function,
which has periodic maximas and minimas. The position of the minimas changes
with the interaction time of atom with the field. Interaction time of the
atom is controlled by controlling the transversal velocity of the atom. Next
atom is sent with different interaction time and eliminates some photon
numbers in the distribution, until after a few number of atoms only one
photon number state is left, which then does not change. By repeating this
simulation and counting the number of times each Fock state is appearing, we
reconstruct the original photon distribution.

In order to apply the procedure we reconstruct the field statistics of a
coherent state developed in the cavity. For an interacting atom undergoing
Bragg scattering, there exists two directions of propagation. Since
undetected atom does not alter the distribution, we can fix our detector at
one of the two directions. The reconstruction of photon statistics is based
on the conditional probability, $P\left( n/p_{l}\left( t\right) \right)
=NP(p_{l}\left( t\right) /n),$ where, $l$ denotes any of the two directions
of propagation of deflected atoms, and $N$ is the normalization constant.
Each atom enters the cavity with a different transversal velocity and,
therefore interacts with the field for a different interaction time. The new
statistics of the cavity field, after the interaction of the atom, updates,
having some photon numbers from the field distribution, eliminated. This way
lack of information about the field reduces with each measurement and the
whole measurement sequence reduces the field to well defined number state,
which then does not change until the field relaxes back to its original
state. We display this procedure in Fig.~3. by displaying a simulation of
eleven such atoms, with a nonrelaxing field initially described by a
coherent state with on average of 10 photons. The effective Rabi frequency
is kept as $\chi =0.02w_{rec}$ to remain within the limit of Bragg regime,
and initial momentum is taken as $P_{0}=4\hbar k,$ in order to have $2nd$
order Bragg scattering. We repeat the process of reduction of field
statistics to a single Fock state and keep record of the number a Fock state
is appearing. This leads us to reconstruct the statistics of the coherent
field state, as we show in Fig.~4 (b). A comparison of original Poisson
statistics with the reconstructed one displays a good agreement in Fig.~4.

We can realize the suggested scheme in laboratory by using the experimental
set up of Ref. \cite{Rempe}. We propagate a beam of rubidium atoms of mass $%
M=1.42\times 10^{-25}$ Kg, through an optical quantum field of wavelength $%
\lambda =0.8$ $\mu m$. Therefore the atom experiences a recoil frequency, $%
w_{rec}=2\pi \times 3.8$ kHz, while passing through the field, provided it
is detuned by an amount $\triangle =2\pi \times 80$ MHz. With these
parameters at hand, we take $g=2\pi \times 112$ kHz, such that, $\chi
\approx 0.02w_{rec},$ in order to realize Bragg regime.

\section{Acknowledgment}

We thank M. S. Zubairy and A. A. Khan for many helpful discussions on the
subject.

\begin{center}
{\bf Figure Captions}
\end{center}

Fig.~1. Suggested experimental setup: We send a highly detuned atomic beam
with momentum component $P_{0}$ along the field, which is very small
compared to the normal component $P_{z}.$ The cavity field of amplitude, $%
{\cal E}($x$)$ is quantized and is along x-axis. In Bragg regime there exist
only two possible directions of deflection for the incident atom, with
momentum $P_{0}$ and $-P_{0}$.

Fig.~2. Numerical verification of higher order Bragg deflection: We study
the occupation probabilities for 2nd order Bragg scattering of atoms
entering with momentum $P_{0}=2\hbar k$ and scattering with momentum (a) $%
P_{0}$ and (b) $P_{-4},$ as a function of interaction time $\bar{t}=w_{rec}t$%
. The probability of finding the atom in either of the two directions flips
between the two states $P_{0}$ and $P_{-4}$ with the same frequency, $B,$ as
obtained by the analytical results of Eqs. $\left( 16\right) $ and $\left(
17\right) $ (dashed line) and by the exact numerical calculation of Eq. $%
\left( 4\right) $ (solid line). All other probabilities are insignificant as
they approach to zero. Here, $\chi =0.02w_{rec}$ and $n=5.$\smallskip

Fig.~3. Reduction of an initial coherent field state to a single photon
number state: In (a), we express initial distribution of coherent field with
average photon number 10. Whereas (b), (c) and (d) display the field
statistics after 2, 6 and 11 detected atoms, respectively. The field
collapses to $n=13$ Fock state. We take the effective Rabi frequency $\chi
=0.02w_{rec},$ as in Fig. 2.

Fig.~4. Reconstruction of photon statistics of the cavity field state: We
display the field initially in coherent state centered at $n=10,$ in Fig.~4
(a), whereas in Fig.~4 (b), we plot the reconstructed field state. The total
number of probe atoms being used is 100,000. The reconstructed statistics is
in good agreement with the original field statistics.


\begin{references}
\bibitem{Ulf}  U. Leonhardt, {\it Measuring the Quantum State of Light }%
(Cambridge University Press, Cambridge, 1997).

\bibitem{tomo}  K. Vogel, and H. Risken, Phys. Rev. A {\bf 40,} 2847 (1989).

\bibitem{Zag}  M. Brune, S. Haroche, V. Lefevre, J. M. Raimond, and N.
Zagury, Phys. Rev. Lett. {\bf 65,} 976 (1990).

\bibitem{endoscopy}  P. J. Bardroff, E. Mayr, and W. P. Schleich, Phys. Rev.
A {\bf 51,} 4963 (1995).

\bibitem{Zub}  M. S. Zubairy, Phys. Lett. A {\bf 222, }91 (1996).

\bibitem{QND}  V. B. Braginsky, Y. I. Vorontsov, and F. Y. Khalili, Zh.
Eksp. Teor. Fiz. {\bf 73,} 1340 (1977).

\bibitem{Kerr}  N. Imoto, H. A. Haus, and Y. Yamamoto, Phys. Rev. A {\bf 32,}
2287 (1985).

\bibitem{Zoller}  M. J. Holland, D. F. Walls, and P. Zoller, Phys. Rev.
Lett. {\bf 67,} 1716 (1991).

\bibitem{Arbab}  A. A. Khan, and M. S. Zubairy, Phys. Lett. A {\bf 254,} 301
(1997).

\bibitem{Mystre}  P. Meystre, E. Schumacher, and S. Stenholm, Opt. Commun. 
{\bf 73,} 443 (1989).

\bibitem{Herkomer}  A. M. Herkommer, V. M. Akulin, and W. P. Schleich, Phys.
Rev. Lett. {\bf 69,} 3298 (1992).

\bibitem{Frey}  M. Freyberger, and A. M. Herkommer, Phys. Rev. Lett. {\bf 72,%
} 1952 (1994).

\bibitem{Zag2}  M. Brune, S. Haroche, J. M. Raimond, L. Davidovich, and N.
Zagury, Phys. Rev. A {\bf 45,} 5193 (1992).

\bibitem{Ueda}  M. Ueda, N. Imoto, H. Nagaoka and T. Ogawa, Phys. Rev.
A.46,. 2859 (1992).

\bibitem{Scully}  M. O. Scully and M. S. Zubairy, Quantum Optics (Cambridge
Univ. Press, Cambridge, 1997).

\bibitem{Shore}  A. F. Bernhardt, and B. W. Shore, Phys. Rev. A {\bf 23,}
1290 (1981).

\bibitem{Marte}  M. Marte, and S. Stenholm, Appl. Phys. B {\bf 54,} 443
(1992).

\bibitem{Braggorg}  W. L. Bragg, Proc. Cambridge Philos. Soc. {\bf 17,} 43
(1912).

\bibitem{Rempe}  S. D$\stackrel{..}{u}$rr, and G. Rempe, Phys. Rev. A {\bf %
59,} 1495 (1999).

\bibitem{Prichard}  P. J. Martin, B. G. Oldaker, A. H. Miklich, and D. E.
Pritchard, Phys. Rev. Lett. {\bf 60,} 515 (1988).

\bibitem{Pichard}  D. E. Pritchard, and P. L. Gould, J. Opt. Soc. Am. B {\bf %
2,} 1799 (1985).

\bibitem{Gould}  P. L. Gould, P. J. Martin, G. A. Ruff, R. E. Stoner, J. L.
Picque, and D. E. Pritchard, Phys. Rev. A {\bf 43,} 585 (1991).

\bibitem{Akulin}  V. M. Akulin, Fam Le Kien, and W. P. Schleich, Phys. Rev.
A {\bf 44,} R1462 (1991).

\bibitem{Leichtle1}  C. Leichtle, W. P. Schleich, I. Sh. Averbukh, and M.
Shapiro, Phys. Rev. Lett. {\bf 80,} 1418 (1998).

\newpage 
\end{references}
\end{document}